\documentclass[letterpaper, 10 pt, conference]{ieeeconf}  
\usepackage{caption}

\IEEEoverridecommandlockouts                              

\usepackage{subcaption}
\usepackage{url}
\usepackage{graphicx}
\usepackage{multirow} 
\usepackage{color}
\usepackage{amsmath,amssymb,amsfonts}
 \usepackage{booktabs} 
\usepackage{threeparttable}

\title{GlaLSTM: A Concurrent LSTM Stream Framework for Glaucoma Detection via Biomarker Relationship Mining}

\author{Cheng Huang, Weizheng Xie, Tsengdar Lee, Karanjit Kooner, Ning Zhang and Jia Zhang\textsuperscript{$\dag$}
\thanks{This research work is partially sponsored by NASA 80NSSC22K0144, UTSW GMO No.241213, and NIH 1R01AG083179-01.}
\thanks{Cheng Huang, Weizheng Xie and Dr. Jia Zhang are with the Department of Computer Science, Southern Methodist University, Dallas, TX 75205, USA. ({\tt\small \{chenghuang, weizhengx, jiazhang\}@smu.edu})}%
\thanks{Dr. Karanjit Kooner is with the Department of Ophthalmology, University of Texas Southwestern Medical Center, Dallas, TX 75390, USA. ({\tt\small karanjit.kooner@utsouthwestern.edu})}%
\thanks{Dr. Tsengdar Lee is with the National Aeronautics and Space Administration (NASA), Washington, DC 20546, USA. ({\tt\small tsengdar.j.lee@nasa.gov})}%
\thanks{Dr. Ning Zhang is with the Department of Pharmacy and Health Systems Sciences, Northeastern University, Boston, MA 02115, USA. ({\tt\small n.zhang@northeastern.edu})}%
\thanks{$\dag$ means corresponding author.}
}

\begin{document}

\maketitle
\thispagestyle{empty}
\pagestyle{empty}

\begin{abstract}

Glaucoma is a leading cause of irreversible blindness globally, commonly linked to elevated intraocular pressure and biomarkers like retinal nerve fiber layer thickness. Understanding how these biomarkers interact is crucial for unraveling glaucoma’s underlying mechanisms. In this paper, we propose GlaLSTM, a novel concurrent LSTM stream framework for glaucoma detection, leveraging latent biomarker relationships. Unlike traditional CNN-based models that primarily detect glaucoma from images, GlaLSTM provides deeper interpretability, revealing the key contributing factors and enhancing model transparency. This approach not only improves detection accuracy but also empowers clinicians with actionable insights, facilitating more informed decision-making. Experimental evaluations confirm that GlaLSTM surpasses existing state-of-the-art methods, demonstrating its potential for both advanced biomarker analysis and reliable glaucoma detection.

\indent \textit{Keywords}— Glaucoma Biomarker, Data Mining, Long short-term memory, Latent Relationship Exploration and Classification
\end{abstract}

\section{INTRODUCTION}

Glaucoma is one of the leading causes of irreversible blindness worldwide, affecting over 76 million people \cite{Harvard-GDP}. Delaying medical attention and treatment can result in blindness, significantly impacting daily life and overall well-being \cite{Kooner-g,G-survey}. However, early detection of glaucoma remains challenging because glaucoma is often asymptomatic in its initial stages, making it difficult to detect until significant vision loss has occurred \cite{LAG}. 

Recently, artificial intelligence (AI) and machine learning (ML) techniques have been increasingly integrated into glaucoma research to enhance early detection \cite{Harvard-GDP,OCTA-500,LAG,xgan}. Given that delayed medical attention and treatment can lead to irreversible blindness, researchers have focused on developing AI-driven solutions to improve diagnostic accuracy \cite{Pieragostino18,Christopher18,Jia14}. Most of these cross-disciplinary efforts are rooted in the domain of computer vision, particularly in image detection and segmentation techniques \cite{Harvard-GDP,OCTA-500,LAG}. Meanwhile, convolutional neural networks (CNNs) \cite{c1,LAG,CNN-1,3d-cnn,xgan} and attention mechanisms \cite{attention,co} have demonstrated impressive performance. However, despite their accuracy, most ML-based models function as ``black boxes," lacking explainability. This opacity raises concerns among clinicians, vision scientists, and patients, who may hesitate to fully trust automated algorithms with unclear decision-making processes.

To bridge this gap, cognitive computing, a branch of AI that replicates human thought processes, offering a promising direction \cite{cc}. Inspired by its principles, we investigates the decision-making approach used by glaucoma clinicians and seeks to translate their diagnostic reasoning into an expert system. By systematically analyzing how clinicians evaluate biomarker relationships in glaucoma detection, we aims to construct a framework that not only enhances classification accuracy but also improves interpretability, making AI-assisted diagnosis more transparent and clinically relevant.

Scrutinizing how clinicians examine patients for glaucoma, we found that their diagnostic approach begins with analyzing a collection of biomarkers \cite{gga,Kooner-g-challenges}, which serve as measurable indicators of the presence, severity, or progression of the disease. In glaucoma assessment, clinicians rely on four primary categories of biomarkers: structural, functional, molecular, and imaging biomarkers. Structural biomarkers include peripapillary retinal nerve fiber layer (RNFL) thickness, optic disc cupping, and ganglion cell complex (GCC) thickness \cite{Huang15}. Functional biomarkers encompass visual field defects \cite{Christopher18} and Electroretinography (ERG) changes \cite{Bach13}. Molecular biomarkers involve proteins in tear fluid or aqueous humor and genetic markers \cite{Pieragostino18}, while imaging biomarkers consist of OCTA-based measurements \cite{Jia14} and fundus photography \cite{Li19}. Since these biomarkers capture different aspects of glaucoma pathology, clinicians integrate them to detect subtle physiological changes that may signal the early onset of the disease.

Inspired by this clinical approach, our research aims to develop an clinical-explainable machine learning framework that mimics how clinicians utilize biomarker relationships for glaucoma detection. Unlike traditional models that rely on black-box parameters \cite{LAG,CNN-1,3d-cnn,c2,c3}, our framework is designed to enhance interpretability by explicitly modeling biomarker interactions via time series models, Long Short-Term Memory Networks (LSTM) \cite{lstm}.  To further refine this approach, multimodal data fusion is employed to enhance biomarker representation, and an attention mechanism \cite{attention,co} is utilized to dynamically assess the relative importance of different biomarkers in glaucoma detection.

All in all, we make the following contributions:
\begin{itemize}

\item We propose a glaucoma detection model, GlaLSTM, based on biomarkers and their interrelationships, which improve diagnostic accuracy and provide insights into the underlying mechanisms of glaucoma.
\item We developed different training strategies based on the diagnostic logic used by doctors for glaucoma detection and conducted targeted comparisons. 
\item Extended experimental results demonstrate that our deep learning framework can effectively detect diseases and accurately identify biomarkers most likely to influence disease progression. 

\end{itemize}

\section{Related Work}

Since glaucoma is a major cause of irreversible blindness worldwide, recent years have witnessed significant AI-empowered research efforts focused on the diagnosis, treatment, and monitoring of glaucoma \cite{Harvard-GDP,OCTA-500,LAG,CNN-1,3d-cnn, Kooner-g-challenges}. And its biomarkers serve as critical indicators for assessing the severity of glaucoma, with intraocular pressure (IOP) being the most significant measure in modern medical evaluations \cite{Kooner-g}. While IOP is the primary indicator, other measures also demonstrate statistical relevance in glaucoma evaluation \cite{Kooner-g-challenges}. Also, non-biomarker factors may also contribute meaningfully. For instance, compared with whites and blacks, Asians are more likely to suffer from angle-closure glaucoma. This is not only related to geography, but also to genes, heredity, physiological structure, etc \cite{G-survey}. 

For the detection task with so many variables mentioned above, study \cite{RNN-LSTM-lp} proposed a deep learning model, RNN-LSTM, which make prediction of the three-month relapse rate for lupus erythematosus rely on analyzing numerous biomarkers alongside relevant patient information, also with non-biomarker factors. Study \cite{cnnlstm} developed a CNN-RNN that not only extracts the spatial features in a fundus image but also the temporal features embedded in a fundus video. However, a limitation of these models are their inability to determine which factors have the greatest impact on the disease. Identifying the most likely causative or aggravating factors in the current circumstances enables doctors to make the most accurate clinical judgments. 

Furthermore, glaucoma is a progressive disease with pathological characteristics that change significantly over time, and these internal factors (biomarkers) are intricately linked \cite{Kooner-g-challenges}. Based on such features, incorporating biomarkers into the model requires accounting for their mutual correlations to ensure accurate analysis. Intraocular pressure (IOP) is the most obvious and intuitive biomarker and many hospitals rely on IOP as the primary step in glaucoma detection \cite{Kooner-g}. Beyond IOP, the influence of multiple biomarkers remains unclear, necessitating a more balanced assessment \cite{Kooner-g-challenges}. Therefore, to minimize data-driven bias, studies implement random confusion groups \cite{XGBoost,Random-Forest,Bi-LSTM} or mixed cross-validation \cite{aensi,sva}, optimizing the balanced contribution of each biomarker. So, aligning with physicians' preferences and the logic of existing medical biomarker testing, a model training strategy incorporating predefined characteristics can more effectively simulate real-life scenarios, enabling accurate disease detection \cite{ConvNet-2020}.

So, in this study, considering the interplay among multiple biomarkers and the temporal progression of glaucoma, we integrated these pathological characteristics to design a deep learning framework, GlaLSTM, capable of classifying glaucoma and identifying the biomarkers most likely to influence disease progression. By identifying the biomarkers with the greatest impact and integrating them with human evaluation, we can provide valuable references for effectively managing the later stages of the disease.

\section{Methodology}
\begin{figure}[t]
  \centering
    \centerline{\includegraphics[width=1.0\columnwidth]{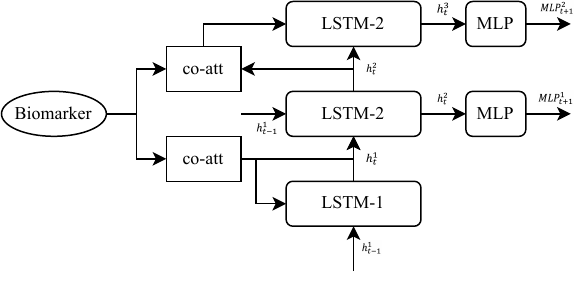}}
  \caption{The architecture of GlaLSTM.}
  \label{stru}
\end{figure}

As illustrated in Fig. \ref{stru}, the concurrent LSTM stream framework comprises two LSTMs and integrates co-attention mechanism \cite{attention,co}, where one LSTM and co-attention are employed twice within the structure. 

While decoding the first co-attention aims to capture features of original input biomarker and its output is $T_{t}^{1}$, sequence and image but the second one focuses on random order input and its output is $T_{t}^{2}$. 

LSTM is a time-series model where earlier inputs gradually lose influence due to the forgetting mechanism, affecting long-term dependencies. We designed a three-stage LSTM framework. Two LSTMs process the same data in different orders to ensure uniform biomarker priority. By comparing their outputs, we identify common and distinct biomarkers. A third LSTM then integrates both sequences, doubling common biomarkers while keeping distinct ones single for better feature extraction.

For the first input of one LSTM, as described in Eq. \ref{eq8}:

\begin{equation}
\label{eq8}
h^{1}_{t} = LSTM(T^{1}_{t},h_{t-1}^{2},W_{e}^{1}x_{t}, h_{t-1}^{1})
\end{equation}

Here, $T^{1}_{t}$ represents the biomarker input to the first LSTM. During the first decoding stage by the other parallel LSTM, the initial weight $W_{e}^{n}x_{t}$ ($n \in \{1,2,3\}$), and the hidden state of the LSTM, $h^{2}_{t-1}$, as given in Eq. \ref{eq9}:

\begin{equation}
\label{eq9}
h_{t}^{2} = LSTM(T_{t}^{2}, h_{t}^{1}, W_{e}^{2}x_{t}, h_{t-1}^{2})
\end{equation}

The outputs of the two LSTMs are passed through independent Multi-Layer Perceptrons (MLPs) to predict the probability distributions of biomarker relationships, as shown in Eq. \ref{eq10}:

\begin{equation}
\begin{gathered}
\label{eq10}
MLP_{t+1}^{1} = softmax(W_{FC1} \times h_{t}^{2})\\
MLP_{t+1}^{2} = softmax(W_{FC2} \times h_{t}^{3})
\end{gathered}
\end{equation}

Where $W_{FC1}$ and $W_{FC2}$ denotes the fully connected layers of the MLP. The third LSTM integrates the outputs of the first two LSTMs, combining their hidden states and weighted features, and the same process applies to the third LSTM, as shown in Eq. \ref{eq11}:

\begin{equation}
\begin{gathered}
h_{t}^{3} = LSTM(h_{t}^{1}, h_{t}^{2}, W_{e}^{3}x_{t})
\label{eq11}
\end{gathered}
\end{equation}

This third LSTM captures higher-order interactions between the biomarker data learned from the first two LSTMs. Features generated by these two MLPs are then utilized to compute the loss functions, which are applied in the subsequent stages of the network. 

For loss function, as shown in Eq. \ref{loss}, $loss1$ ($loss2$) is the cross-entropy loss. They are used to evaluate the errors caused by the two LSTMs used to calculate the input. Since the third LSTM is based on the input of the first two, the evaluation significance of it is not as direct as the first two. From the beginning, we set $\lambda\in(0,1]$) to 0.5 at first. we set its coefficient $\alpha\in[1,10)$ at 5. Different parameter settings are used to evaluate the error changes that can be produced by two parallel LSTMs when facing the same batch of data sets but with different input orders.

\begin{equation}
\label{loss}
l o s s=\lambda*l o s s_{1}+\alpha*l o s s_{2}
\end{equation}

\section{Data and its Pre-processing}

We have two sources of data. One is a public source\footnote{\url{https://huggingface.co/datasets/AswanthCManoj/glaucoma_diagnosis_json_analysis}} and the other is from University of Texas Southwestern Medical Center (UTSW). UTSW’s institutional review board (IRB) approved this study, which followed the principles of the Declaration of Helsinki. Since the study was retrospective, the IRB waived the requirement for informed consent from patients.

\begin{table}[h]
\caption{One Sample from UTSW}
\begin{center}
\begin{tabular}{c|c|c|c|c}
\toprule
\multirow{2.5}{*}{\textbf{Category}} & \multirow{2.5}{*}{\textbf{\textbf{Biomarker}}}& \multicolumn{3}{c}{\textbf{Detail}}\\
\cmidrule{3-5} 
 & & \textbf{\textit{OD}}& \textbf{\textit{OS}} & \textbf{IE (OD-OS)}\\
\midrule
\multirow{4}{*}{RNFL} & AR (µm) & 97 & 91 &6\\
& SR (µm) & 94 & 92 &2\\
& IR (µm) & 99 & 89 &10\\
& I-ER (S-I) (µm) & -5& 3&N/A\\
\midrule
\multirow{6}{*}{ONH} & A-O & 0.28 & 0.51 &-0.23\\
& V-O & 0.46 & 0.79 &-0.33\\
& H-O & 0.62 & 0.74 &-0.12\\
& RA (mm$^{2}$) & 1.44 & 1.23 &0.21\\
& DA (mm$^{2}$) & 2.01 & 2.52 &-0.51\\
& CVO (mm$^{2}$) & 0.043 & 0.299 &-0.256\\
\midrule
\multirow{6}{*}{GCC} & A-G (µm) & 85 & 83 &2\\
& S-G (µm) & 81 & 83 &-2\\
& I-F (µm) & 88 & 83 &5\\
& I-EG (S-I) (µm) & -7 & 0 &N/A\\
& FLV  & 0.87 & 0.47 & 0.40\\
& GLV  & 10.76 & 12.50 &-1.74\\
\bottomrule
\end{tabular}
\begin{threeparttable}
 \begin{tablenotes}
        \footnotesize
        \item $^{1}$retinal nerve fiber layer = RNFL (Average RNFL = AR; Superior RNFL = SR; Inferior RNFL = IR; Intra Eye (S-I) = I-ER) 
        \item $^{2}$optic nerve head = ONH (Cup/Disc Area Ratio = A-O; Cup/Disc V. Ratio = V-O; Cup/Disc H. Ratio = H-O; Rim Area = RA; Disc Area = DA; Cup Volume = CVO)
        \item $^{3}$ganglion cell complex = GCC (Average GCC = A-G; Superior GCC = S-G; Inferior FCC = I-F; Intra Eye (S-I) = I-EG (S-I))
        \item $^{4}$IE (OD-OS) is a condition that affects both the right and left eyes.
        \item $^{5}$right eye = OD and left eye = OS
        \item $^{6}$Except for GCC, RNFL and ONH, the remaining abbreviations are only used in this paper.
        \item $^{7}$The number of training and test samples is 500 and 100 respectively.
        \item $^{8}$IOP is an additional input.
      \end{tablenotes}
  \end{threeparttable}
\label{data-utsw}
\end{center}
\end{table}

For private data from UTSW, as shown in Table. \ref{data-utsw}, this is a sample collected from a healthy person. For its preprocessing, there are three types of biomarkers themselves: Within Normal, Borderline and Outside Normal. For instance, A-G is the borderline and its label is ``Borderline" or 0.5, S-G is outside normal, labeled ``False" or 0,  and AR is nomal, labeled ``True" or 1. Among them, Borderline is a critical situation, which is normal in medical terms, but its value is close to the pathology. Unless explicitly stated as normal, it is classified into a third category. In addition, we also considered some other non-biomarker influencing factors: age, gender, race, region, etc. Finally, we have two glaucoma experts from UTSW in our team who can help us with annotations. 

\begin{table}[h]
\caption{One Sample from Public Dataset}
\begin{center}
\begin{tabular}{c|c}
\toprule
\textbf{Fundus\_Features} & \textbf{Ground True} \\
\midrule
optic\_disc\_size  & large \\
cup\_to\_disc\_ratio  & 0.8 \\
neuroretinal\_rim  & thinned  \\
isnt\_rule\_followed  & false \\
rim\_pallor  & true \\
rim\_color  & pale \\
bayoneting  & true \\
sharp\_edge  & true \\
laminar\_dot\_sign  & true \\
notching  & true \\
rim\_thinning  &  true \\
additional\_observations  & null  \\
\midrule
\textbf{glaucoma\_risk\_assessment}  &  \textbf{high risk}\\
\textbf{confidence\_level}  & \textbf{0.9 }\\
\bottomrule
\end{tabular}
\label{data-public}
  \begin{threeparttable}
 \begin{tablenotes}
        \footnotesize
        \item $^{1}$``neuroretinal\_rim" has been simplified and the original version is thinned in all regions, especially inferior and superior."
        \item $^{2}$``glaucoma\_risk\_assessment" and``confidence\_level" are not biomarkers. They are pre-judged by humans to predict the probability of glaucoma and their own confidence in that judgment.
        \item $^{3}$The annotation of this sample is ``glaucoma".
        \item $^{4}$The number of training and test samples is 589 and 100 respectively.
      \end{tablenotes}
  \end{threeparttable}
\end{center}
\end{table} 

For public data, as shown in Table. \ref{data-public}, compared with data in the Table. \ref{data-utsw}, the amount of data has been reduced significantly. For its preprocessing to  features like optic\_disc\_size, cup\_to\_disc\_ratio and neuroretinal\_rim which have specific value or description, we will refer to the judgment criteria for glaucoma and give ``True" or ``False". Unlike Table. \ref{data-utsw}, glaucoma\_risk\_assessment and confidence\_level are indicators that are prejudged, which will greatly affect the potential relationship mining of various biomarkers. Moreover, the OS and OD parameters, representing the left and right eyes, along with the IE (OD-OS) parameter in the data from Table. \ref{data-utsw}, are three times the number of parameter names listed, whose variable value (32*3, single eye; 48*3, double eyes) is 6 or 4 times greater than those (12*2) in Table. \ref{data-public} (in this case, ``Borderline" is reguarded as nomal, or the third category, it is 4 or 2.7 times). In this study, the experiment we designed for UTSW is for both eyes.

\section{Experimental Result and Analysis}

\subsection{Classification Results}

\begin{table}[t]
\caption{Experimental Results on Public Data}
\begin{center}
\begin{tabular}{c|c|c|c|c}
\toprule
\textbf{Model} & \textbf{Prejudge} & \textbf{Precision} & \textbf{Recall} & \textbf{F1-score}\\
\midrule
\multirow{2}{*}{RNN-LSTM \cite{RNN-LSTM-lp}}  
   & $\times$       & 0.723 & 0.731 & 0.719 \\
& $\sqrt{}\mkern-9mu{\smallsetminus}$       & 0.683 & 0.711 & 0.694 \\
   & \checkmark     & 0.847 & 0.801 & 0.824 \\
\midrule
\multirow{2}{*}{Random Forest \cite{Random-Forest}}  
   & $\times$       & 0.713 & 0.737 & 0.724 \\
& $\sqrt{}\mkern-9mu{\smallsetminus}$       & 0.693 & 0.724 & 0.717 \\
   & \checkmark     & 0.837 & 0.816 & 0.831 \\
\midrule 
\multirow{2}{*}{Bi-LSTM \cite{Bi-LSTM}}  
   & $\times$       & 0.724 & 0.745 & 0.733 \\
& $\sqrt{}\mkern-9mu{\smallsetminus}$       & 0.711 & 0.733 & 0.715 \\
   & \checkmark     & 0.842 & 0.812 & 0.827 \\
\midrule
\multirow{2}{*}{Decision Tree \cite{PrivaTree}}  
   & $\times$       & 0.767 & 0.783 & 0.769 \\
& $\sqrt{}\mkern-9mu{\smallsetminus}$       & 0.667 & 0.685 & 0.673 \\
   & \checkmark     & 0.818 & 0.786 & 0.811 \\
\midrule
\multirow{2}{*}{XGBoost \cite{XGBFEMF}}  
   & $\times$       & 0.739 & 0.736 & 0.740 \\
& $\sqrt{}\mkern-9mu{\smallsetminus}$       & 0.737 & 0.733 & 0.739 \\
   & \checkmark     & 0.881 & 0.868 & 0.879 \\
\midrule
\multirow{2}{*}{Ours}  
   & $\times$       & 0.756 & 0.762 & 0.753 \\
& $\sqrt{}\mkern-9mu{\smallsetminus}$      & 0.733 & 0.751 & 0.745 \\
   & \checkmark     & 0.903 & 0.917 & 0.913 \\
\midrule
\multirow{2}{*}{GPT-4o \cite{GPT4}}  
   & $\times$       & 0.740 & 0.730 & 0.735 \\
& $\sqrt{}\mkern-9mu{\smallsetminus}$       & 0.726 & 0.717 & 0.729  \\
   & \checkmark     & 0.891 & 0.869 & 0.878 \\
\bottomrule
\end{tabular}
\begin{threeparttable}
\begin{tablenotes}
    \footnotesize
    \item $^{1}$if glaucoma\_risk\_assessment and confidence\_level are enabled, giving them a higher weight, the thief is \checkmark; otherwise, the weight is the same as other weights, then $\sqrt{}\mkern-9mu{\smallsetminus}$; if not enabled, then $\times$.
\end{tablenotes}
\end{threeparttable}
\label{er-c-2}
\end{center}
\end{table}

As shown in Table. \ref{er-c-2}, it presents the experimental results of various models evaluated on public data, comparing their performance in terms of Precision, Recall, and F1-score, both with and without the ``Prejudge" mechanism. Enabling Prejudge consistently improves all models' performance. For instance, RNN-LSTM’s F1-score rises from 0.694 without Prejudge to 0.824 with full Prejudge, and similar improvements are observed for Random Forest (0.717 to 0.831) and Bi-LSTM (0.715 to 0.827). Notably, Decision Tree and XGBoost also show substantial gains in F1-score with full Prejudge, reaching 0.811 and 0.879, respectively. Among the models, XGBoost demonstrates more stable performance metrics compared to GPT-4o, although the overall results of the two are roughly comparable. Our proposed model achieves the highest overall performance, with an F1-score of 0.913 under full Prejudge, highlighting its superior ability to handle classification tasks. 

\begin{table}[ht]
\caption{Experimental Results on Data from UTSW}
\tiny
\begin{center}
\begin{tabular}{c|c|c|c|c|c|c}
\toprule
\textbf{Model} & \textbf{Random} & \textbf{Borderline} & \textbf{Others} & \textbf{Precision} & \textbf{Recall} & \textbf{F1-score}\\
\midrule
\multirow{8}{*}{RNN-LSTM \cite{RNN-LSTM-lp}}  
  & $\times$ & $\times$ & $\times$       & 0.720 & 0.690 & 0.705 \\
  & $\times$ & $\times$ & \checkmark     & 0.680 & 0.620 & 0.649 \\
  & $\times$ & \checkmark & $\times$     & 0.650 & 0.590 & 0.618 \\
  & $\times$ & \checkmark & \checkmark   & 0.610 & 0.550 & 0.578 \\
  & \checkmark & $\times$ & $\times$     & 0.590 & 0.530 & 0.558 \\
  & \checkmark & $\times$ & \checkmark   & 0.570 & 0.510 & 0.538 \\
  & \checkmark & \checkmark & $\times$   & 0.550 & 0.500 & 0.524 \\
  & \checkmark & \checkmark & \checkmark & 0.530 & 0.490 & 0.509 \\
\midrule
\multirow{8}{*}{Random Forest \cite{Random-Forest}}  
  & $\times$ & $\times$ & $\times$       & 0.780 & 0.760 & 0.770 \\
  & $\times$ & $\times$ & \checkmark     & 0.750 & 0.720 & 0.735 \\
  & $\times$ & \checkmark & $\times$     & 0.720 & 0.690 & 0.705 \\
  & $\times$ & \checkmark & \checkmark   & 0.690 & 0.650 & 0.669 \\
  & \checkmark & $\times$ & $\times$     & 0.670 & 0.620 & 0.644 \\
  & \checkmark & $\times$ & \checkmark   & 0.640 & 0.590 & 0.614 \\
  & \checkmark & \checkmark & $\times$   & 0.610 & 0.560 & 0.584 \\
  & \checkmark & \checkmark & \checkmark & 0.600 & 0.540 & 0.569 \\
\midrule
\multirow{8}{*}{Bi-LSTM \cite{Bi-LSTM}}  
  & $\times$ & $\times$ & $\times$       & 0.779 & 0.741 & 0.759 \\
  & $\times$ & $\times$ & \checkmark     & 0.738 & 0.692 & 0.713 \\
  & $\times$ & \checkmark & $\times$     & 0.711 & 0.649 & 0.679 \\
  & $\times$ & \checkmark & \checkmark   & 0.672 & 0.592 & 0.631 \\
  & \checkmark & $\times$ & $\times$     & 0.659 & 0.581 & 0.619 \\
  & \checkmark & $\times$ & \checkmark   & 0.619 & 0.531 & 0.571 \\
  & \checkmark & \checkmark & $\times$   & 0.589 & 0.519 & 0.553 \\
  & \checkmark & \checkmark & \checkmark & 0.601 & 0.501 & 0.549 \\
\midrule
\multirow{8}{*}{Decision Tree \cite{PrivaTree}}  
  & $\times$ & $\times$ & $\times$       & 0.650 & 0.670 & 0.659 \\
  & $\times$ & $\times$ & \checkmark     & 0.661 & 0.690 & 0.675 \\
  & $\times$ & \checkmark & $\times$     & 0.730 & 0.750 & 0.740 \\
  & $\times$ & \checkmark & \checkmark   & 0.760 & 0.780 & 0.770 \\
  & \checkmark & $\times$ & $\times$     & 0.670 & 0.680 & 0.675 \\
  & \checkmark & $\times$ & \checkmark   & 0.691 & 0.710 & 0.700 \\
  & \checkmark & \checkmark & $\times$   & 0.770 & 0.790 & 0.780 \\
  & \checkmark & \checkmark & \checkmark & 0.800 & 0.790 & 0.795 \\
\midrule
\multirow{8}{*}{XGBoost \cite{XGBFEMF}}  
  & $\times$ & $\times$ & $\times$       & 0.721 & 0.751 & 0.734 \\
  & $\times$ & $\times$ & \checkmark     & 0.768 & 0.732 & 0.749 \\
  & $\times$ & \checkmark & $\times$     & 0.794 & 0.812 & 0.803 \\
  & $\times$ & \checkmark & \checkmark   & 0.832 & 0.771 & 0.800 \\
  & \checkmark & $\times$ & $\times$     & 0.751 & 0.783 & 0.766 \\
  & \checkmark & $\times$ & \checkmark   & 0.782 & 0.763 & 0.772 \\
  & \checkmark & \checkmark & $\times$   & 0.811 & 0.832 & 0.821 \\
  & \checkmark & \checkmark & \checkmark & 0.834 & 0.801 & 0.819 \\
\midrule
\multirow{8}{*}{Ours}  
  & $\times$ & $\times$ & $\times$       & 0.781 & 0.792 & 0.786 \\
  & $\times$ & $\times$ & \checkmark     & 0.803 & 0.811 & 0.807 \\
  & $\times$ & \checkmark & $\times$     & 0.824 & 0.837 & 0.830 \\
  & $\times$ & \checkmark & \checkmark   & 0.842 & 0.851 & 0.846 \\
  & \checkmark & $\times$ & $\times$     & 0.802 & 0.814 & 0.808 \\
  & \checkmark & $\times$ & \checkmark   & 0.823 & 0.832 & 0.827 \\
  & \checkmark & \checkmark & $\times$   & 0.851 & 0.861 & 0.856 \\
  & \checkmark & \checkmark & \checkmark & 0.862 & 0.853 & 0.857 \\
\midrule
\multirow{8}{*}{GPT-4o \cite{GPT4}}  
  & $\times$ & $\times$ & $\times$       & 0.731 & 0.742 & 0.734 \\
  & $\times$ & $\times$ & \checkmark     & 0.753 & 0.768 & 0.761 \\
  & $\times$ & \checkmark & $\times$     & 0.774 & 0.783 & 0.778 \\
  & $\times$ & \checkmark & \checkmark   & 0.792 & 0.801 & 0.796 \\
  & \checkmark & $\times$ & $\times$     & 0.751 & 0.764 & 0.758 \\
  & \checkmark & $\times$ & \checkmark   & 0.773 & 0.782 & 0.779 \\
  & \checkmark & \checkmark & $\times$   & 0.813 & 0.821 & 0.817 \\
  & \checkmark & \checkmark & \checkmark & 0.832 & 0.823 & 0.827 \\
\bottomrule
\end{tabular}
\begin{threeparttable}
\begin{tablenotes}
    \footnotesize
    \item $^{1}$``Random" means that ignoring the order of IOP, all biomarkers were initially randomized ($\times$) or on the contrary, IOP presets the highest priority (\checkmark).
    \item $^{2}$``Borderline" means that biomarkers with label `Borderline' are uniformly classified as the third category (\checkmark), otherwise they are classified as the positive category ($\times$). 
    \item $^{3}$``Others" means that if some of the non-biological marker information mentioned earlier, such as age, gender, etc., is taken into account (\checkmark), otherwise not ($\times$). 
\end{tablenotes}
\end{threeparttable}
\label{er-c-1}
\end{center}
\end{table}

As shown in Table \ref{er-c-1}, the concurrent LSTM stream framework achieves the highest F1-score (0.857) with all conditions enabled, demonstrating its robustness. XGBoost (0.819) and GPT-4o (0.827) also show strong, stable performance. In contrast, RNN-LSTM, Random Forest, and Bi-LSTM perform better with fewer conditions but decline under full settings, indicating sensitivity to complexity and noise. Decision Tree performs consistently, peaking at 0.795 with all conditions. These results highlight the proposed model’s effectiveness and the stability of XGBoost and GPT-4o in diverse settings.

\begin{figure*}[h]
  \centering
  \begin{subfigure}{0.245\linewidth}
    \centerline{\includegraphics[width=\columnwidth]{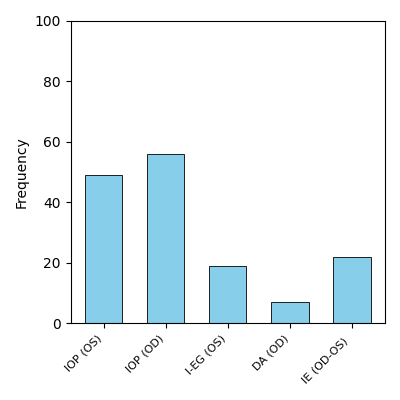}}
    \caption{$\times$ $\times$ $\times$}
    \label{000}
  \end{subfigure}
  \hfill
  \begin{subfigure}{0.245\linewidth}
\centerline{\includegraphics[width=\columnwidth]{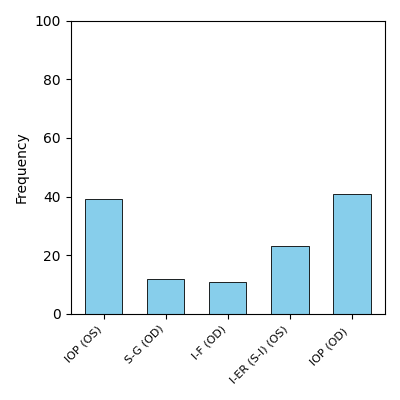}}
    \caption{$\times$ $\times$ \checkmark}
    \label{001}
  \end{subfigure}
    \hfill
  \begin{subfigure}{0.245\linewidth}
\centerline{\includegraphics[width=\columnwidth]{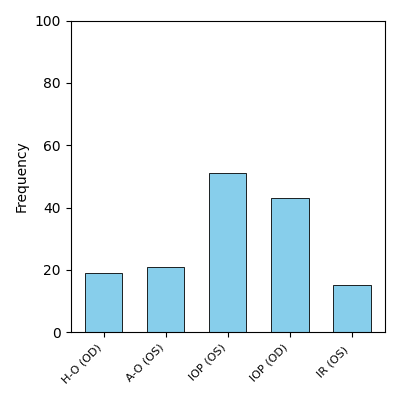}}
    \caption{$\times$ \checkmark $\times$}
    \label{010}
  \end{subfigure}
    \hfill
  \begin{subfigure}{0.245\linewidth}
\centerline{\includegraphics[width=\columnwidth]{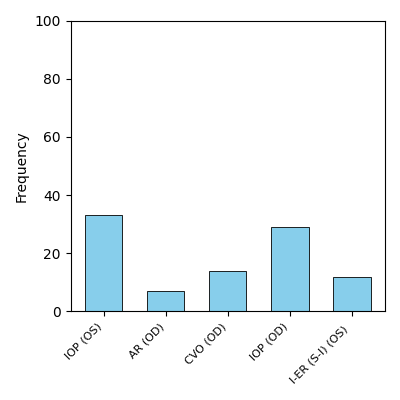}}
    \caption{$\times$ \checkmark $\times$}
    \label{011}
  \end{subfigure}
    \hfill
  \begin{subfigure}{0.245\linewidth}
\centerline{\includegraphics[width=\columnwidth]{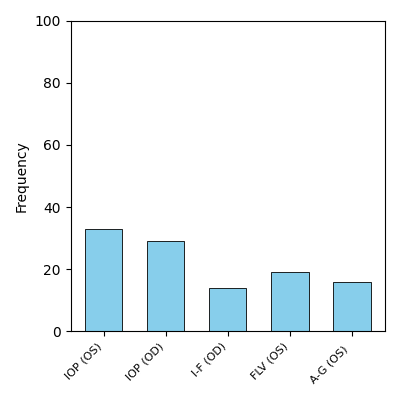}}
    \caption{\checkmark $\times$ $\times$}
    \label{100}
  \end{subfigure}
    \hfill
  \begin{subfigure}{0.245\linewidth}
\centerline{\includegraphics[width=\columnwidth]{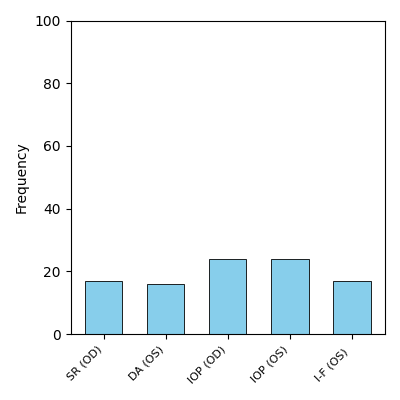}}
    \caption{\checkmark $\times$ \checkmark}
    \label{101}
  \end{subfigure}
    \hfill
  \begin{subfigure}{0.245\linewidth}
\centerline{\includegraphics[width=\columnwidth]{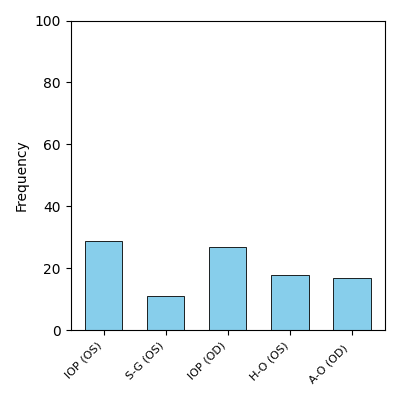}}
    \caption{\checkmark \checkmark $\times$}
    \label{110}
  \end{subfigure}
    \hfill
  \begin{subfigure}{0.245\linewidth}
\centerline{\includegraphics[width=\columnwidth]{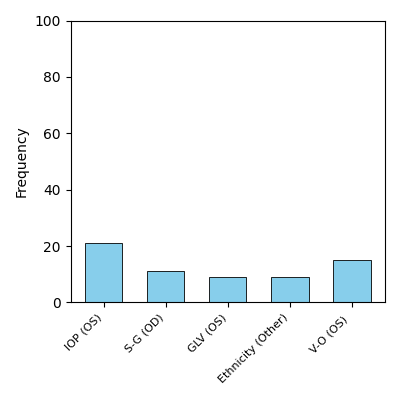}}
    \caption{\checkmark \checkmark \checkmark}
    \label{111}
  \end{subfigure}
  \caption{The following statistics pertain to the five most influential biomarkers (UTSW), displayed horizontally in descending order of influence from left to right. Amony them, the symbols below each image are for the three categories in Table. \ref{er-c-1}: ``Random", ``Borderline" and ``Others".}
  \label{UTSW-bio}
\end{figure*}

\subsection{Hidden Relationship Mining}

As shown in Fig. \ref{UTSW-bio}, the vertical axis is set to 100, representing 100 test groups. IOP is a key indicator for glaucoma diagnosis, consistently ranking in the top five regardless of data standard divisions. This aligns with clinical practice, where IOP is the first metric assessed \cite{Kooner-g}.

\begin{figure}[h]
  \centering
  \begin{subfigure}{0.49\linewidth}
    \centerline{\includegraphics[width=\columnwidth]{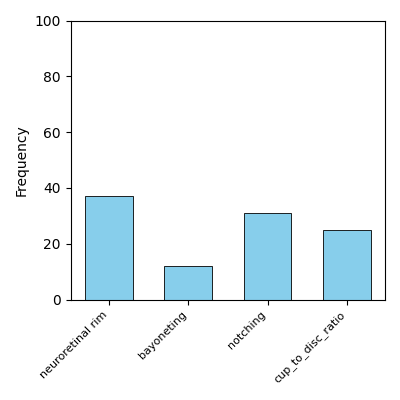}}
    \caption{$\times$}
    \label{000}
  \end{subfigure}
  \hfill
  \begin{subfigure}{0.49\linewidth}
\centerline{\includegraphics[width=\columnwidth]{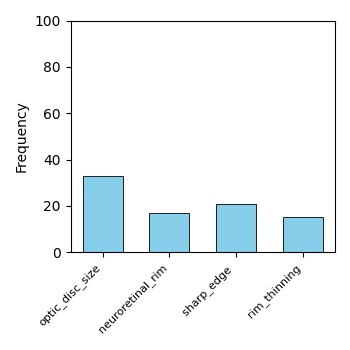}}
    \caption{\checkmark}
    \label{001}
  \end{subfigure}
  \caption{The following statistics pertain to the five most influential biomarkers (public data), displayed horizontally in descending order of influence from left to right. Amony them, the symbol below each image are for the category in Table. \ref{er-c-2}: "Prejudge". } 
  \label{pd-bio}
\end{figure}

\subsection{Analysis}

Our analysis reveals that as the number of calculated variables increases, the model must balance more factors, resulting in lower accuracy when the number of training iterations remains constant. In contrast, if the model is deliberately given a training bias by assigning a higher initial weight to ``Prejudge (\checkmark)," the accuracy of the models improves significantly, as shown in Table. \ref{er-c-2}. These results demonstrate the significant impact of Prejudge on improving model accuracy and emphasize the effectiveness of the proposed framework. Additionally, we observed that when the ``Prejudge" condition is assigned the same weight as other biomarkers, the performance of all models declines in the experimental results.

Table \ref{er-c-1} Analysis: Enabling the Borderline indicator led to a significant performance drop for RNN-LSTM, Random Forest, and Bi-LSTM, while improving other models. This decline stems from overfitting tendencies, Random Forest struggles with extrapolation, and LSTMs require more data to retain distant sample features. In contrast, decision trees excel due to intrinsic feature selection and multi-type data handling but are highly sensitive to data fluctuations, causing result instability. XGBoost mitigates this via iterative boosting, regularization, and automatic tree pruning, enhancing stability and performance. Unlike decision trees, XGBoost also optimally handles missing values, making it a more robust choice.

Overall, compared to Table \ref{er-c-2}, the values in Table \ref{er-c-1} are generally lower. The key factor is Borderline, when set to ``$\times$", the magnitude increases only slightly per case, regardless of being ``True" or ``False." However, when set to \checkmark, it effectively introduces a new major category, neutral samples. Treating the process as a multivariate function, this addition corresponds to a new dimension, significantly increasing computational complexity by an entire order of magnitude. Concurrent LSTM stream framework excelled across all metrics, especially after data shuffling, enhancing generalization and accuracy. Using two LSTMs as head inputs broadens feature learning, preventing local optima. Different input orders improve sampling, reducing overfitting. Compared to GPT-4o’s Prompt Project results, our approach is more stable and noise-resistant.

Also, several details are noteworthy for classification: Fig. \ref{111} shows that ethnicity under the "Others" category appears once, reflecting glaucoma's racial differences, consistent with existing literature \cite{Kooner-g,Kooner-g-challenges,G-survey}. Additionally, when ``Borderline" changes to $\times$, the model integrates more variables, balancing IOP’s significance by introducing additional factors. While elevated IOP is a primary glaucoma marker, other elements—such as genetics, lifestyle, and regional history—also play crucial roles. Thus, despite IOP's high frequency in the experiment, the underlying contributing factors warrant deeper consideration.

\section{Conclusions}

In this paper, we introduce GlaLSTM, a novel concurrent LSTM stream framework designed for glaucoma detection through biomarker relationship mining. Unlike traditional machine learning models that function as black boxes, GlaLSTM enhances interpretability by explicitly modeling the intrinsic relationships among glaucoma biomarkers, enabling more clinically transparent decision-making. Our framework effectively integrates biomarkers, mimicking the diagnostic reasoning of clinicians. Through extensive experiments on both public and clinical datasets, GlaLSTM demonstrates superior performance in classification accuracy, recall, and F1-score, outperforming state-of-the-art approaches such as CNN-RNN and Transformer-based methods. Furthermore, our ablation studies confirm the effectiveness of biomarker relationship analysis in improving diagnostic reliability. These findings highlight the clinical significance of GlaLSTM as a trustworthy, interpretable, and data-driven tool for early glaucoma detection and progression analysis. Future work will explore multi-modal data fusion, real-time clinical deployment, and extended validation across diverse populations to further enhance the applicability of our framework in ophthalmology and beyond.

\end{document}